\documentclass[12pt]{article}
\usepackage[english]{babel}
\usepackage{amssymb}
\usepackage{amssymb,amsmath,amsfonts}
\usepackage{mathtools}

\newcommand\btheorem{\vspace{0.31em}\par\refstepcounter{thrm}\bf T\,h\,e\,o\,r\,e\,m\,~\thethrm\,.\,~\sl}
\newcommand\etheorem{\rm\vspace{0.3em}\par}
\newcommand\proofr{{\it Proof.\,~}}
\def\proofend{$\Box$\vspace{0.3em}\par}

\title {   On non-full-rank perfect codes \\ over finite fields }
\author {Alexander~M.~Romanov\thanks{Sobolev Institute of Mathematics, Russian Academy of Sciences,  4 Academician Koptyug avenue, 630090 Novosibirsk, Russia.  Email: rom@math.nsc.ru}
}
\date{}

\begin{document}

\maketitle

\abstract{The paper deals with the  perfect 1-error correcting codes  over a finite field with $q$ elements (briefly  $q$-ary 1-perfect codes).
We show that the orthogonal  code to the $q$-ary non-full-rank 1-perfect code of length $n = (q^{m}-1)/(q-1)$ is
a $q$-ary constant-weight code with  Hamming weight equals to $q^{m - 1}$ where $m$ is any natural number not less than two.
We derive necessary and sufficient conditions for  $q$-ary  1-perfect codes of non-full rank.
We  suggest a generalization of the concatenation construction to  the $q$-ary case and
construct the ternary 1-perfect codes of length 13 and rank 12.}

\section{Introduction}\label{sec:intr}

Let $\mathbb{F}_{q}^n$ be a vector space of dimension $n$ over the finite field $\mathbb{F}_{q}$, where  $q = p^r$,
$p$ is a prime number,  $r$ is a positive integer.
The \emph{Hamming distance} between two vectors, ${\bf x}$, ${\bf y} \in \mathbb{F}_{q}^n$, is the number of coordinates in which they differ, denoted by $d({\bf x}, {\bf y})$.
An arbitrary subset  ${C}$ of  $\mathbb{F}_{q}^n$ is called a \emph{$q$-ary  1-perfect}   code of length $n$ if for every vector ${\bf x} \in \mathbb{F}_{q}^n$ there exists a unique vector ${\bf c} \in  {C}$ such that
$d({\bf x},{\bf c}) \leq 1$. Non-trivial $q$-ary 1-perfect  codes of length $n$   exist only
if $n = (q ^{m} -1) / (q-1)$, where $m$ is a natural number not less than two.
Two codes ${C}_1, {C}_2 \subseteq \mathbb{F}_{q}^n $ are said to be \emph{equivalent} if there exists a vector
${\bf v} \in \mathbb{F}_{q}^n $ and $n \times n$ monomial matrix $M$ over $\mathbb{F}_{q}$ such that
${C}_2 =\{({\bf v} + {\bf c}M) \, \,  |  \, \,  {\bf c} \in  {C}_1\}$.
We assume that the all-zero vector ${\bf 0}$ is in code. A code is called \emph{linear} if it is a linear space over $\mathbb{F}_{q}$. A linear $q$-ary  1-perfect code of length $n$ is unique up to equivalence and is called a \emph{$q$-ary Hamming code}.
Nonlinear $q$-ary 1-perfect codes exist for  $q = 2$, $m \geq 4$, $q \geq 3$, $m \geq 3$, $q \geq 5$, $m \geq 2$.
For $q =4 $, $m = 2$, the question of the existence of nonlinear $q$-ary $1$-perfect codes still remains open,
see \cite{rom7, lin, ph5, ph51}.

The \emph{rank} of code $C$ is the maximum number of linearly independent codewords of $C$.  A code of length $n$ that has  rank $n$ is said to have \emph{full rank}; otherwise, the code is \emph{non-full rank}.
The rank of $q$-ary  Hamming code of length $n$ is equal to $n - m$ where $n = (q^m - 1) / (q - 1)$, $m \geq 2$.

The switching constructions of  full-rank  $q$-ary   1-perfect codes have been proposed for all
$n = (q^m - 1) / (q - 1)$, where $m \geq 4$, see \cite{etz, rom, ph5}.
For $m = 3$ and for  $q = p^r$, $r > 1$, the existence  of full-rank  $q$-ary $1$-perfect codes is proved in \cite{ph51}.
The question of the existence of full-rank  $q$-ary $1$-perfect codes of length $n = (q^m - 1) / (q - 1)$
still remains  open if $m = 3$, $q \geq 3$, $q$ is a prime number, and if $m = 2$, $q \geq 4$, see \cite{ph5, ph51}.

In the $q$-ary case, the question of the minimum and maximum possible cardinality of the intersection of two distinct  $q$-ary $1$-perfect codes of the same length is still open. In the binary case,  this question was
answered in \cite{etz, etz2}.

It is established  that there exist at least $q^{q^{cn}}$  nonequivalent $q$-ary  1-perfect codes of length $n$, where $c = \frac{1}{q} - \epsilon $, see \cite{ vas1, sch, lin}.

A \emph{concatenation construction} of 1-perfect codes is a combinatorial generalization of the well-known
$ (\bf u | \bf u + \bf v) $ construction. We give a description of the concatenation construction for the binary case.

Let $m$ be any natural number not less than two and $n = 2^{m} - 1$.
Next, let $C_{0}^1, C_{1}^1, \ldots, C_{n}^1$ be a partition of the vector space $\mathbb {F}_{2}^{n}$ into binary
1-perfect codes of length $n$ and let $C_{0}^2, C_{1}^2, \ldots, C_{n}^2$ be a partition of the binary trivial MDS code with parameters $[n + 1, n, 2]$ into binary extended 1-perfect codes with parameters $(n + 1, 2^{n - m}, 4)$.
(A trivial MDS code with parameters $[n + 1, n, 2]$ consists of all binary vectors of length $n + 1$ of even weight.)
Then the given partitions $C_{0}^1, C_{1}^1, \ldots, C_{n}^1$, $C_{0}^2, C_{1}^2, \ldots, C_{n }^2$ and
the permutation $\alpha$ acting on the index set  $I = \{0, 1, \ldots, n \}$ determine the binary 1-perfect code

$$C_{\alpha} = \{({\bf u}| {\bf v}) \, \,  |  \, \, {\bf u} \in C_{i}^1 , {\bf v} \in C_{\alpha(i)}^2   \}$$
of length  $2n + 1$.

In the binary case, the concatenation construction is based on partitions of two types --  partitions of the space  $\mathbb {F}_{2}^{n}$ into 1-perfect codes and partitions  of the binary trivial MDS code into extended 1-perfect codes.
For many decades, the question of the parameters of codes to which we must partition the trivial MDS codes in the
$q$ -ary case remained open.
In this paper we obtain an answer to this question and generalize the concatenation construction to the $q$-ary case.

For $m \geq 4$, $q$-ary 1-perfect codes of length $n = (q^m-1) / (q-1)$ and rank $n-m + s$ exist for all
$s \in \{0, 1, \ldots, m \}$, see \cite{etz, rom, ph5}.
For $m = 3$ and for  $q = p^r$, $r > 1$, the existence of $q$-ary 1-perfect  codes of length
$n = (q^3 - 1) / (q - 1)$ and  rank $n - 3 + s$, $s \in \{1, 2, 3 \}$ is proved in \cite {ph51}.
For $m = 3$, $q \geq 3$,  Lindstr\"{o}m--Sch\"{o}nheim  codes are codes of length $n = (q^ 3 - 1) / (q - 1)$ and rank $n - 2$, see \cite{sch, lin}.
For $m = 3$, $q \geq 3$, and for $q$, which is a prime number, the question of the existence of $q$-ary 1-perfect
codes of length $n = (q ^ 3 - 1) / (q - 1)$ and rank $n - s$, $s \in \{1, 0 \}$ still remains open \cite{ph5, ph51}.
In particular, the question of the existence of ternary 1-perfect codes of length 13 and rank 12 remained open.
In this paper we have an answer to this question. Using the $q$-ary concatenation construction proposed in this paper, we construct the ternary 1-perfect codes of length 13 and rank 12.

In Section~\ref{sec:orth}, we show that the code  orthogonal to the  non-full-rank $q$-ary 1-perfect code of length
$n = (q^{m} -1) / (q-1)$ is a $q$-ary constant-weight code with Hamming weight  equals
to $q^{m - 1}$, where $m$ is any natural number not less than two.
In Section~\ref{sec:neces}, we obtain necessary and sufficient conditions for $q$-ary 1-perfect codes of non-full-rank.
In Section~\ref{sec:con}, we  suggest a generalization of the concatenation construction to the $q$-ary case.
In Section~\ref{sec:ter}, by the concatenation construction, ternary 1-perfect codes of length 13 and rank 12 are constructed.

\section{Orthogonal codes}\label{sec:orth}
Let $ {\bf u}, {\bf v} \in \mathbb {F}_{q}^{n}$. Then the \emph{scalar product} of the vectors
${\bf u} = (u_1, u_2, \ldots, u_n)$ and ${\bf v} = (v_1, v_2, \ldots, v_n) $ is the  mapping
$${\bf u} \cdot {\bf v} = \sum_{i = 1}^n u_iv_i.$$
Let $C \subset \mathbb{F}_{q}^{n}$. Then the  code
$$C^{\bot} = \{ {\bf u}  \, \,  |  \, \, {\bf u} \in \mathbb{F}_{q}^{n}  \,\,  \mbox{and}  \,\, {\bf u}\cdot{\bf v} = 0
\mbox{ for all} \,   {\bf v} \in C \}$$

is  \emph{orthogonal}  or  \emph{dual}  of code $C$.
The dual code of Hamming code is simplex code.
The simplex code has parameters $[n,m,q^{m-1}]_q$, $n = (q^{m}-1)/(q-1)$.
All words of a simplex code, with the exception of the  all-zero  word, have the same weight equal to  $q^{m -1}$.

Following \cite{etz} we will use the methods described in \cite{mac}, page 132.
We shall represent sets of vectors from the space $\mathbb{F}_{q}^{n}$ by formal polynomials in
variables $z_1, z_2, \ldots, z_n$.
The vector  ${\bf v} = (v_1,v_2, \ldots, v_n) \in \mathbb{F}_{q}^{n}$ corresponds to
monomial ${\bf z}^{\bf v}  = {z_1}^{v_1} {z_2}^{v_2}  \cdots {z_n}^{v_n}$.
The set of all monomials in the variables $z_1, z_2, \ldots, z_n$ forms a multiplicative group $G$ whose multiplication corresponds to the addition in $\mathbb{F}_{q}^{n}$.
Next, we define the group algebra $QG$ of the group $G$ over the field of rational numbers $Q$, which consists of all formal sums
$$\sum_{{\bf v} \in \mathbb{F}_{q}^{n}}a_{\bf v}{\bf z}^{\bf v}, {\mbox{ãäå}} \, \, \,  a_{\bf v} \in Q, \, \, {\bf z}^{\bf v} \in G.$$
For each ${\bf u} \in \mathbb {F}_{q}^{n}$, we define the character of the group $G$.
Let $\chi_{\bf u}({\bf z}^{\bf v})  = \zeta^{{{\bf u}\cdot{\bf v}}}$,
where $\zeta$ is a primitive  $q$th root  of unity. The character $\chi_{\bf u}$ is extended  on $QG$ by linearity.
Then
$$ \chi_{\bf u}(C)  = \chi_{\bf u}\left(\sum_{{\bf v} \in C}{{\bf z}^{\bf v}}\right) =
\sum_{{\bf v} \in C}\chi_{\bf u}({{\bf z}^{\bf v}}) =
\sum_{{\bf v} \in C}\zeta^{{\bf u}\cdot{\bf v}}.$$
It is known that if the code $C$ is linear, then

\[ \chi_{\bf u}(C)  = \left\{ \begin{array}{ccl}
 |C|& {\mbox{if}} &  {\bf u} \in C^{\bot}, \\
  0 & {\mbox{if}} &  {\bf u} \notin C^{\bot}.
 \end{array}\right. \]

\btheorem
\label{th1}
Let $C$ be a $q$-ary 1-perfect code of length $n = (q^{m} - 1) / (q - 1)$ and rank $k$.
Let $\{B_0, B_1, \ldots, B_n \}$ be the weight distribution  of the code $C^{\bot}$.
Then
\[ B_i =  \left\{ \begin{array}{lcrcl}
 1 &            \mbox{if} &                       &i& = 0,  \\
 0 &            \mbox{if} &                1 \leq &i&\leq  q^{m - 1}  - 1,\\
 q^{n-k} - 1 &  \mbox{if} &                       &i& = q^{m - 1},  \\
 0 &            \mbox{if} &   q^{m - 1}  + 1 \leq &i& \leq n.
\end{array} \right. \]

\etheorem
\proofr
Let $V \subset \mathbb{F}_{q}^{n}$ be a Hamming sphere of radius 1.
Since the code $C$ is $q$-ary 1-perfect, then $\mathbb{F}_{q}^{n} = C + V$ and, consequently,
$\chi_{\bf u}(\mathbb{F}_{q}^{n}) = \chi_{\bf u}(C)\chi_{\bf u}(V) = 0.$
Thus $\chi_{\bf u}(C) = 0$  if $\chi_{\bf u}(V) \neq 0$.
Since $\sum_{i = 0}^{q - 1}\zeta^i = 0$ è $\zeta^0 = 1$, then $\chi_{\bf u}(V) = q^m - q{\cdot}wt({\bf u)} \neq 0$
for $wt({\bf u)} \neq q^{m -1}$ (the Hamming weight  of the vector ${\bf u}$ is denoted by $wt({\bf u)}$).
\proofend

\section{Necessary and sufficient conditions}\label{sec:neces}

We denote by $M_{q, n}$ the $q$-ary trivial MDS code  (\emph{maximum distance separable code})
with parameters $[n,n-1,2]_q$.
Consider a code $C$ of length $n = (q^{m}-1)/(q-1)$ over the field $\mathbb{F}_{q}$.
Let a vector  $({\bf u}|{\bf v}) \in C$.
We assume that $\bf u$ has length equal to $n - q^{m - 1}$.
For each vector ${\bf v} \in M_{q, q^{m-1}}$, we define the $q$-ary code $C'({\bf v})$ of length $n - q^{m - 1}$.
Let
$$ C'({\bf v}) = \{{\bf u} \in  \mathbb{F}_{q}^{n -  q^{m - 1}} \, \,  |  \, \, ({\bf u}|{\bf v}) \in C  \}. $$
For each vector ${\bf u} \in  \mathbb{F}_{q}^{n -  q^{m - 1}}$, we define the $q$-ary code $C''({\bf u})$
of length $q^{m - 1}$. Let
$$ C''({\bf u}) = \{{\bf v} \in M_{q,q^{m - 1}} \, \,  |  \, \, ({\bf u}|{\bf v}) \in C  \}. $$
\btheorem
\label{th2}
For $q \neq 2$, the code $C$ of length $n = (q^{m} -1) / (q-1)$ over the field $\mathbb{F}_{q}$ is
a $q$-ary 1-perfect code of non-full-rank if and only if,
when the code $C'({\bf v})$ is a $q$-ary 1-perfect code of length $n - q^{m -1}$
for any ${\bf v} \in M_{q, q^{m-1}}$
and the code $C''({\bf u}) \subset M_{q,q^{m - 1}}$ is a $q$-ary code with parameters $(q^{m-1}, q^ {q^{m-1} - m}, 3)_q$
for any $ {\bf u} \in \mathbb {F}_{q}^{n - q ^ {m-1}}$.
\etheorem
\proofr
Let us prove the sufficiency of the conditions of the theorem.
For this it is necessary to show that the number of code words in the code $C$ is correct and the minimum distance
$d(C)$ of the code $C$ is 3.
Let us show that the number of code words in the code $C$ is correct.
By the definition of the code $C'$, we have

$$ |C| = \sum_{{\bf v} \in M_{q,q^{m - 1}} } |C'({\bf v})| = q^{q^{m - 1} - 1}q^{n - q^{m - 1} - (m - 1)}
= q^{n - m}.$$

By the definition of the code $C''$, we have

$$ |C| = \sum_{ {\bf u} \in \mathbb{F}_{q}^{n -  q^{m - 1}}} |C''({\bf u})| =
q^{n - q^{m - 1} } q^{q^{m - 1} - m} = q^{n - m}.$$

Now we show that the minimum distance $d(C) = 3$.
Let the words $({\bf u}|{\bf v})$ è $({\bf u}'|{\bf v}')$ belong to the code $C$.
Assume that ${\bf u} = {\bf u}'$, then $ {\bf v}, {\bf v}' \in  C'' $ and, therefore,
$d(({\bf u}|{\bf v}),({\bf u}'|{\bf v}')) \geq 3.$
Assume that ${\bf v} = {\bf v}'$, then $ {\bf u}, {\bf u}' \in  C' $  and
$d(({\bf u}|{\bf v}),({\bf u}'|{\bf v}')) \geq 3.$
Further assume that ${\bf u} \neq {\bf u}'$ è ${\bf v} \neq {\bf v}'$, then  $d({\bf u},{\bf u}') \geq 1$.  Since ${\bf v},{\bf v}' \in  M_{q,q^{m - 1}}$, then $d({\bf v},{\bf v}') \geq 2$ and, therefore,
$d(({\bf u}|{\bf v}),({\bf u}'|{\bf v}')) \geq 3$.

Consider the vector  ${\bf w} \in \mathbb{F}_{q}^{n}$.
Suppose that the first $n - q^{m - 1}$ components of the vector $\bf w$ are equal to 0, and the remaining components of this vector are equal to 1.
Then it is obvious that $\bf w \in C^{\bot}$ and, therefore, the code $C$ is a $q$-ary 1-perfect code of non-full-rank.

Next we prove the necessity of the conditions of the theorem.
By the conditions of the theorem, the code $C$ is a $q$ -ary 1-perfect code of non-full rank and, consequently, there exists a nonzero vector $\bf w \in C^{\bot}$.
Then it follows from  Theorem \ref{th1} that $wt({\bf w}) = q^{m-1}$.
Suppose that the first $n - q^{m - 1}$ components of the vector $\bf w$ are equal to 0, and the remaining components of this vector are equal to 1.
Consider the vector $({\bf u}|{\bf v}) \in C$.
Assume that  $\bf u$ has a length equal to $n - q^{m - 1}$.
Then, since the first $n - q^{m - 1}$ components of the vector $\bf w$ are equal to 0,
then ${\bf v} \in M_{q,q^{m - 1}}$.

Since  $d(C) \geq 3$, then $d(C'({\bf v})) \geq 3$.
Since the length of the code $C'({\bf v})$ is equal to $n - q^{m - 1}$, it follows from the sphere-packing bound  that $|C'({\bf v})| \leq q^{n - q^{m - 1} - (m -1)}$.
Assume that
$|C'({\bf v'})| < q^{n - q^{m - 1} - (m -1)}$  for some ${\bf v'}$. Then

$$ |C| = \sum_{{\bf v} \in M_{q,q^{m - 1}} } |C'({\bf v})| = (q^{q^{m - 1} - 1} - 1)q^{n - q^{m - 1} - (m - 1)}
+ |C'({\bf v'})| < q^{n - m},$$
which contradicts the condition of the theorem on the perfectness of the code $C$.

Since the code $C$ is a $q$-ary 1-perfect code of length $n$ and the length of the code $C''({\bf u})$
is equal to $q^{m-1}$, it is known that $|C''({\bf u})| \leq q^{q^{m - 1} - m}$  see  \cite{ mac, del}.
Assume that
$|C''({\bf u'})| < q^{q^{m - 1} - m}$ for some ${\bf u'}$.  Then
$$ |C| = \sum_{{\bf u} \in  \mathbb{F}_{q}^{n -  q^{m - 1}}}
|C''({\bf u})| = ({q}^{n -  q^{m - 1}} - 1)q^{q^{m - 1} - m} + |C''({\bf u'})| < q^{n - m}.$$
\proofend

For $q = 2$, the code $C''(\bf u)$ has the parameters $(2^{m - 1},2^{2^{m - 1} - m},4)$, see \cite{etz}.

From Theorem 2.1 of the paper \cite{hed2} it follows, in particular, that for any $q$-ary 1-perfect code  $C$ of non-full rank and length $n = (q^{m} -1) / (q-1)$ there exists a monomial transformation $\psi$ of the space $\mathbb{F}_{q}^{n}$ such that

\begin{equation}
\psi(C) =\{({\bf u}|{\bf v}) \, \,  |  \, \, {\bf u} \in  C'({\bf v}), {\bf v} \in M_{q,q^{m - 1}} \},
\end{equation}
where $C'({\bf v})$ is a $q$-ary 1-perfect code of  length ${n -  q^{m - 1}}$.

It follows from the  Theorem \ref {th2} that any vector ${\bf w} \in C^{\bot}$ forms a representation of the code
$C$ in the form (1).

\section{Concatenation construction}\label{sec:con}

Let $q \neq 2$, $m$ be any natural number not less than two, and $n = (q^{m} -1) / (q-1)$.
Next, let $C_{0}^1 C_{1}^1, \ldots, C_{(q - 1)n}^1 $  be a partition of the vector space $\mathbb{F}_{q}^{n}$ into
$q$-ary 1-perfect codes of length $n$, and let $C_{0}^2, C_{1}^2, \ldots, C_{(q - 1)n}^2 $ be a partition of $q$-ary trivial MDS code with parameters  $[(q - 1)n + 1, (q - 1)n, 2]_q$ into $q$-ary codes with parameters
$((q - 1)n + 1, q^{(q - 1)n - m}, 3)_q$
(such a partition exists by  Theorem  \ref{th2}).

\btheorem
\label{th3}
Given  partitions
$C_{0}^1, C_{1}^1, \ldots, C_{(q - 1)n}^1 $,   $C_{0}^2, C_{1}^2, \ldots, C_{(q - 1)n}^2$, and the permutation $\alpha$ acting on the index set $ I = \{0, 1, \ldots , (q - 1)n\}$ determine the $q$-ary 1-perfect code

$$C_{\alpha} = \{({\bf u}| {\bf v}) \, \,  |  \, \, {\bf u} \in C_{i}^1 , {\bf v} \in C_{\alpha(i)}^2   \}$$
of length $qn + 1$.
\etheorem
\proofr
We need to show that the number of code words in the code $C_{\alpha}$ is correct and the minimum distance
$d(C_{\alpha})$ of the code $C_{\alpha}$ is 3.
Since a $q$-ary 1-perfect code of length $n$ contains $q^{n-m}$ codewords, then

$$ |C_{\alpha}| = \sum_{{\bf v} \in M_{q,(q - 1)n + 1} } |C'({\bf v})| = q^{{(q - 1)n} }q^{n - m}
= q^{qn + 1 - (m + 1)}.$$

Next we show that the minimum distance $ d(C_{\alpha})$ = 3.
Let the words $({\bf u}|{\bf v})$ è $({\bf u}'|{\bf v}')$  belong to the code $C_{\alpha}$.
Assume that  ${\bf v} = {\bf v}'$, then $ {\bf u}, {\bf u}' \in  C_{i}^1 $
for some  $i \in I$ and, consequently,
$d(({\bf u}|{\bf v}),({\bf u}'|{\bf v}')) \geq 3.$
Assume that ${\bf u} = {\bf u}'$, then $ {\bf v}, {\bf v}' \in  C_{\alpha(i)}^2 $ and
$d(({\bf u}|{\bf v}),({\bf u}'|{\bf v}')) \geq 3.$
Further assume that ${\bf u} \neq {\bf u}'$ è ${\bf v} \neq {\bf v}'$, then  $d({\bf u},{\bf u}') \geq 1$.
Since ${\bf v},{\bf v}' \in  M_{q,(q - 1)n + 1}$, then $d({\bf v},{\bf v}') \geq 2$   and, therefore,
$d(({\bf u}|{\bf v}),({\bf u}'|{\bf v}')) \geq 3$.
\proofend

By Theorem \ref {th2} (for the binary case, see. \cite{etz}) concatenation construction does not allow to construct  full-rank codes, but allows constructing codes belonging to different switching classes  \cite{ph2}.
(It is known that codes of full rank can be constructed using the method of switching
for all $n = (q ^ {s} -1) / (q-1)$, where $s \geq 4$, see \cite {etz, rom, ph5}.)
Etzion and Vardy \cite{etz} showed that by concatenation method all binary 1-perfect codes of non-full rank of length $n$ can not be constructed.
They proposed the construction of binary 1-perfect codes, which is based on the so-called \emph{"perfect segmentations}" and showed that their method allows obtaining codes  different from binary 1-perfect codes constructed by the concatenation method.
Heden \cite{hed1} based on the partitioning of the space $\mathbb{F}_{2}^{7}$ into binary Hamming codes and partitions of the binary trivial MDS code of length 8 into binary extended Hamming codes using the concatenation construction constructed a binary 1-perfect code of length 15, which is not the code of Vasil'ev \cite{vas1}.
Solov'va \cite{sol} investigated the properties of binary 1-perfect codes, which are constructed by the concatenation method from partitions of the space $\mathbb {F}_{2}^{n}$ into Vasil'ev codes \cite{vas1} and from the partition of MDS codes into extended  Vasil'ev codes.
There are many papers in which various partitions of the space $\mathbb{F}_{q}^{n}$ into 1-perfect codes are proposed (see, for example, \cite{krot}).
Phelps \cite{ph1} proposed a concatenation construction of binary extended 1-perfect codes of length $2n + 2$ starting from partitions of the binary trivial MDS code with parameters $[n + 1, n, 2]$ into binary extended 1-perfect codes of length $n + 1$.
Phelps \cite{ph3} enumerated all pairwise nonequivalent binary extended 1-perfect codes of length 16,
which can be constructed by concatenation, and
showed that this method allows  to construct at least 963 such codes.
As is known \cite{ost2}, there are 2165 pairwise nonequivalent binary extended 1-perfect codes of length 16.
The number of pairwise nonequivalent binary extended 1-perfect codes of length 16 and rank 15 is equal to 175.
Therefore, we can assume that by concatenation it is impossible to construct all binary extended 1-perfect codes of length $16$, whose rank is less than 15.
Phelps \cite{ph4} generalized the binary concatenation construction that doubled the length of the code, and instead of permutations suggested using quasigroups; herewith  the length of the code began to increase many times.
Heden and Krotov \cite{hed2} generalized the Phelps construction \cite{ph4} to the $q$-ary case.
The construction of Heden and Krotov is based on partitions  of the space $\mathbb{F}_{q}^{n}$ into $q$-ary 1-perfect codes and codes with parameters $((q - 1) n + 1, q^{(q - 1) n-m}, 3)_q$ are not used in it.

\section{Ternary codes}\label{sec:ter}
The ternary Hamming code of length 13 has rank 10.
The ternary nonlinear 1-perfect Lindstr\"{o}m--Sch\"{o}nheim  codes of length 13 have a rank equal to 11, see \cite{sch, lin}.
In this section, using the concatenation method, we construct a ternary 1-perfect code of length 13 and rank 12.

Let the matrix
\[\left[ \begin{array}{ccccccccccccc}
0 & 0 & 0 & 0 & 1 & 1 & 1 & 1 & 1 & 1 & 1 & 1 & 1 \\
0 & 1 & 1 & 1 & 0 & 0 & 0 & 1 & 1 & 1 & 2 & 2 & 2 \\
1 & 0 & 1 & 2 & 0 & 1 & 2 & 0 & 1 & 2 & 0 & 1 & 2 \\
\end{array}\right]\]
be a parity-check matrix  of the ternary Hamming  code of length 13, which we denote by $H$.
Consider the vector $0000111111111 \in H^{\bot}$.
According to the theorem \ref{th2}, this vector generates a ternary linear code $C''$ with parameters $[9,6,3]_3$, whose codewords belong to the ternary trivial MDS code $M_{3,9}$ with parameters $[9,8,2]_3$.
Since the code $C''$ is linear and $C'' \subset M_{3,9}$, it generates a partition of $M_{3.9}$ into cosets.
The number of  cosets is 9.
Let $m = 2$. Then the ternary Hamming code of length 4 forms a partition of the space $\mathbb{F}_{3}^{4}$ into cosets. The number of cosets is also 9.
A ternary 1-perfect code of length 13 and rank 12, constructed using the concatenation construction, we denote
by $C_\alpha$.
The permutation $\alpha$ is chosen in such a way that the code $C_\alpha$ contains the all-zero  vector.
It is obvious that for some permutation $\alpha$ the vectors

\[ \begin{array}{ccccccccccccc}
0 & 0 & 0 & 1 & 1 & 2 & 0 & 0 & 0 & 0 & 0 & 0 & 0 \\
0 & 0 & 0 & 2 & 1 & 0 & 2 & 0 & 0 & 0 & 0 & 0 & 0 \\
0 & 0 & 1 & 0 & 1 & 0 & 0 & 2 & 0 & 0 & 0 & 0 & 0 \\
0 & 0 & 2 & 0 & 1 & 0 & 0 & 0 & 2 & 0 & 0 & 0 & 0 \\
0 & 1 & 0 & 0 & 1 & 0 & 0 & 0 & 0 & 2 & 0 & 0 & 0 \\
0 & 2 & 0 & 0 & 1 & 0 & 0 & 0 & 0 & 0 & 2 & 0 & 0 \\
1 & 0 & 0 & 0 & 1 & 0 & 0 & 0 & 0 & 0 & 0 & 2 & 0 \\
2 & 0 & 0 & 0 & 1 & 0 & 0 & 0 & 0 & 0 & 0 & 0 & 2 \\
0 & 0 & 0 & 0 & 0 & 0 & 0 & 1 & 1 & 1 & 0 & 0 & 0 \\
0 & 0 & 0 & 0 & 0 & 0 & 1 & 0 & 1 & 0 & 1 & 0 & 0 \\
0 & 0 & 0 & 0 & 0 & 1 & 0 & 0 & 1 & 0 & 0 & 1 & 0 \\
0 & 0 & 0 & 0 & 0 & 0 & 0 & 0 & 1 & 2 & 0 & 2 & 1 \\
\end{array}\]
belong to the code $C_\alpha$ and are linearly independent.

The question of the existence of ternary 1-perfect codes of length 13 and rank 13 still remains open.

 \end{document}